# Controlling Citizen's Cyber Viewing Using Enhanced Internet Content Filters

**Shafi'í Muhammad ABDULHAMID** (MIEEE, MNCS, NAICSIT, MIAENG)
Cyber Security Science Department, Federal University of Technology Minna, Nigeria
*E-mail: shafii.abdulhamid@futminna.edu.ng*

**Fasilat Folagbayo IBRAHIM** (MNACOSS, MNAMS)
Mathematics and Statistics Department, Federal University of Technology Minna, Nigeria
*E-mail: fasilat100@gmail.com*

*Abstract*— Information passing through internet is generally unrestricted and uncontrollable and a good web content filter acts very much like a sieve. This paper looks at how citizens' internet viewing can be controlled using content filters to prevent access to illegal sites and malicious contents in Nigeria. Primary data were obtained by administering 100 questionnaires. The data was analyzed by a software package called Statistical Package for Social Sciences (SPSS). The result of the study shows that 66.4% of the respondents agreed that the internet is been abused and the abuse can be controlled by the use of content filters. The PHP, MySQL and Apache were used to design a content filter program. It was recommended that a lot still need to be done by public organizations, academic institutions, government and its agencies especially the Economic and Financial Crime Commission (EFCC) in Nigeria to control the internet abuse by the under aged and criminals.

## I. Introduction

The internet delivers cyber surfers with unrestricted information and opportunities to connect and share information across the world. Data passing through the cyberspace are generally open and uncontrollable. The internet users can access and share images, audios and videos of violence, pornography, material about prohibited drugs, computer viruses and malwares through the cyberspace. There are several websites and applications hosting and sharing adult-oriented resources. According to the figures reported in [1], 230 million pornographic web pages access are being filtered out of thirty nine billion through their content filters. The issues become even more complex when the cyberspace atrocities involve a state of not being known or identified by name (anonymity), publication considered bellicose or a menace to the security and reorganize in this way. In the traditional media such as television programs and motion pictures, legislation can lessen the problems. Nonetheless, in the global and the domestic network, the person that is not barred from ethical principles can basically hold their resources in a nation with less restrictive bylaws and policies. Consequently, the uncoordinated regulations cannot function effectively at the national level. To protect children and checkmate grownups from retrieving prohibited materials from the cyberspace while at work in the offices, content grading schemes and blocking applications have been designed by software houses or companies [2].

It appears that those persons whose standpoint was for strict sieving and those who have a contrary opinion and supported uncontrolled and complete liberty for cyber content access are all not correct. No doubts, the access to cyberspace should be moderated through filtering and regulations. Though, the practice of using cyber content filters should be sensible in order to offer the young ones an opportunity to feel the atmosphere of information age and, at the same time, to allow the young ones not to face cyber contents that are mostly not destined for children. For example, broadcasting violence, explicit sex contents, drugs, etc. [3].

The conception of this notion is not as easy as it might appear at a glimpse. Although the issue of censorship of cyber contents did not misplace its actuality, the purposes of this issue do exist. The supporters of the use of cyber filters claim that citizens deserve the best possible opportunities and at the same time, should be protected from viewing the illegal content. Yet, the problems of societal implications and the effectiveness of filtering and internet blocking programs become the issues of the day. With rapid growth of internet, it becomes more and more difficult to determine the content of internet sites that should be legitimately required to be blocked.

A Web filter is a program that can screen an incoming Web page to determine whether some or all of it should not be displayed to the user. The filter checks the origin or content of a Web page against a set of rules provided by company or person who has installed the Web filter [4]. With the advent of Web 2.0 technologies, websites are now mash-ups of content that is aggregated from many other sites. This scenario adds complexity to filtering websites based on domain names alone and also opens up new avenues of attack for hackers and virus writers who are becoming increasingly successful at compromising syndicated feeds. If just one feed of data is compromised, all the websites that pull in that feed will deliver malicious code to their trusted users. An effective content filtering solution will judge incoming web data based on its content and not its source alone. Malicious content that is smuggled into trusted sites will still be detected and filtered out thereby protecting the internal network.

In case a user needs access to specific internet site that is currently blacklisted, they can apply to authorities to get access to the website. The opponents of use of internet filters predominantly claim that blocking and filtering the content of web sites violate human right but the basic argument is that the use of internet filters is ineffective, unnecessary and discriminatory. This research work seeks to check and control unauthorized and fraudulent sites using content filters.

The aim of this paper is to focus on how to control the cyber access of the citizens using an enhanced internet content filter. Section II discusses the related works. Section III explains the different types of content filters. Section IV explains the materials and method used and the last section, which is section V is for the conclusions and recommendations.

## II. Related Works

Katherine [5] is of the opinion that individuals are using the cyberspace more often to make new friends, find love and to begin excessive affairs and provided a careful appraisal of the literature on cyber infidelity. The research further explained that the normal age of a kid when first exposed to cyber pornography is eleven years old, with the highest surfers of pornography sites in the United States of America being the 12 to 17-year old range. Nearly 90 percent of eight to 16-year-olds has watched pornography through the cyberspace while doing assignments at homework. Of those who were accidentally exposed to pornography while surfing the Web, 66 percent say they did not seek the images out and did not want to view them.

Barbara *et al.* [6] said that the disorder of primary societies and opted to metropolitan life, with its attendant loss of rich cultural values has encroached on the adolescent's ability to handle their newly awakened sexual impulses. These authors noticed gross sexual misconducts among different age groups in our nation by the claim of urbanization, modernization, spurious sexual expressions in junk magazines and of course pornography and internet dating. The consequences are not farfetched; they include child-pregnancy, abortion, sexually transmitted diseases and of course possible increase in the incidence of Acquired Immune Deficiency Syndrome (AIDS) due to unguided sexual escapades.

Across all the tribes represented in Nigeria, there are forefathers who gave young ones informal/traditional education to prepare them for life. These traditional education systems from the North to the South and the East to the West maintained that young men and women were taught to acquire a healthy attitude towards sex as part of their preparation for adult life. The expressions of sexuality were delayed until the youngsters are matured enough to face the challenges of adulthood [7].

Research shows that in Nigeria, over 40% of Internet usage is related to browsing of sex sites. The remaining 60% is distributed among searches for information on academics, entertainment, migration, sports and etc. [8]. In continuation to the research [9], reviews the impact of Internet pornography on web-users in Nigeria and advocates the use of web filtering programs as a robust measure against unwanted Internet content.

The opponents of use of internet filters predominantly claim that blocking and filtering the content of web sites violate human rights. The basic argument is that the use of internet filters is ineffective, unnecessary and discriminatory; it demeans the status of teachers as well as students, and violates constitutional rights [10]. According to them, when one group of people or a person (e.g. a Board of Education, a superintendent, a school director, etc.) decides what the other people should think and read, they are exceeding their proper role and probably violating the First Amendment of the U.S. Constitution [11].

According to the research study conducted by the Phillips [12], the use of internet filters Bess (developed by N2H2 and Surf Control companies) erroneously block the vast majority of quite legitimate educational web sites simply because of the blacklist of key words that contains words that are relatively admissible. In another research paper, the regulatory option of internet filtering measures was put into the broader perspective of the legal framework regulating the (exemption from) liability of Internet Service Providers (ISPs) for user-generated contents. In addition, the paper suggests proposals on which regulatory decisions can better ensure the respect of freedoms and the protection of rights. The paper introduces several significant cases of blocking online copyright infringing materials. Copyright related blocking techniques have been devised for business reasons – by copyright holders' associations. It must be recalled, however, that these blocking actions cannot be enforced without the states' intervention [13].

Of all of the innumerable myths that swarm around cyberspace, one of the most insidious is that the internet is an inherently emancipatory tool, a device that

    

necessarily and inevitably promotes democracy by giving voice to those who lack political power, and in so doing undermines authoritarian and repressive governments [14]. But as the size of global computer networks expands and the use of the Internet skyrockets, the security issues do manifest themselves not only in the security of computer networks but also in individual user security on individual PCs connected to the Internet either via an organization's gateway or an Internet service provider (ISP). The security of every user, therefore, is paramount whether the user is a member of an organization network or a user of a home PC via an independent ISP. In either case, the effort is focused on protecting not only the data but also the user [15 -16]. In another research, a Gabor filter was applied for fiducial point localization [17]. Methods, devices, and products provide for restricting access to mature content by individuals for whom access to the mature content is designated as inappropriate. A content filter receives a communication, determines that the communication includes an image, and extracts the image. The image is scanned for mature content. A content restrictor component restricts access by various classes of users to the mature content [18-19].

In another research paper, an interface was developed that can help parents, teachers and kids to search the web in a secure way and make the interaction session with the web a pleasurable experience. The Algorithm discussed will keep the kid focused by not blocking or filtering the websites but by redirecting the interest of the kid towards the educational side of his interest. The interface designed using the algorithm will not block any website but has redirected the query to the educational part of kids' interest [20, 21, 22].

## III.  Types of Content Filters

Content filters can be classified into the following five broad categories;

### 3.1  Browser based Filters:

Browser based content filtering solution is the most lightweight solution to do the content filtering in which there is no need to install native software or library to home computer unlike Google Chromium and Firefox browser in which most of their browser add-on/extensions are written in JavaScript which must reviewed and tested by a review editor team before releasing to internet users. The major advantage of the browser based content filtering solution is that add-on/extension solution is green, clean and easy to install and uninstall.

### 3.2  Client – Side Filters:

Client – Side filters is installed directly on computers. This type of filters uses a password and only those with the password can change the filters settings. Users can customize client side filters to meet their specific needs and is best used for homes or businesses that need to filter only certain computers or sets of computers.

### 3.3  Content-Limited ISPs:

Content-Limited ISPs are filters created by Internet Service Provider which are not limited to specific users or computers but to everyone who uses the service. In addition to filtering or blocking certain sites, many Content-Limited ISPs also monitor e-mail and chat traffic, blocking inappropriate users or sites as needed and prevent any chat or e-mail message from anyone except other users of the ISPs.

### 3.4  Search-Engine Filters:

Search-Engine Filters helps users to filters out inappropriate material on a search engine level such as Google or Yahoo. Users can turn on the filters and this filter does not prevent a user from visiting a website with inappropriate content but does not display the sites in search results. It is useful when trying to avoid certain viruses or pornography that may use misleading description to entice you to visit.

### 3.5  Server-Side Filters:

Server-Side Filters affect every users on the network in which businesses or institutions receive access from these filters since they can set one filter for all computers and users. It is installed on a central server computer which is connected to other computers on the same network.

## IV.  Materials and Method

The survey method employed was the use of questionnaire, which solicits information from respondents selected for the research. The questionnaire titled *"Internet Content Filters and Internet Abuse"* was administered to respondents in different locations in Nigerian. A total of 100 questionnaires were administered over a one-month period of which only 86 were returned. In other to get a more reliable data set, confidentiality of personal information provided by the respondents was guaranteed and some internet terms were explained to assist the respondents in understanding each question. The primary data generated were analyzed statistically by Statistical Package for Social Sciences (SPSS) and Microsoft Excel. Fig. 1 below shows the summary of responses of the raw data obtained from the questionnaire.

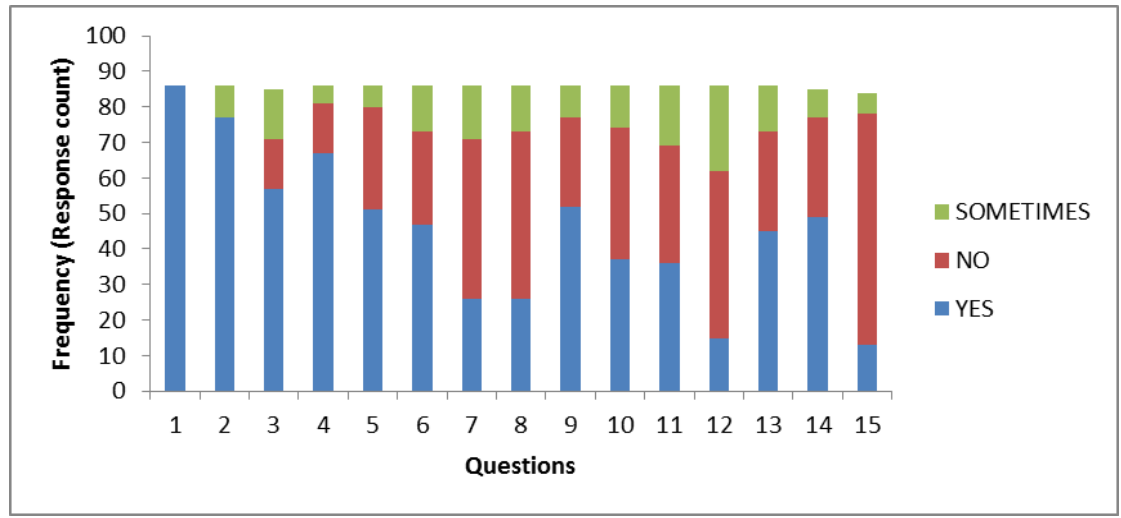

Fig. 1: A component bar chart illustrating a summary of responses (Questions 1 – 15 as indicated in Table 1)

## 4.1  Data Analysis

The results of the Chi-Square (χ-Square) analyses are presented in this section in other to provide a richer understanding of the target audience's perception of this study. Table 1 summarizes the frequencies and corresponding percentages for the general public perceptions to this study. It reveals that 100.0% of the respondents know what the internet is, but only 89.5% of these respondents use the internet and about 10.5% of the total respondents sometimes use the internet.

Table 1: summary of the frequencies and corresponding percentages of respondents' perceptions of using content filter

| QUESTION | FREQUENCY | | | PERCENTAGE | | | MEAN | STANDARD DEVIATION |
|---|---|---|---|---|---|---|---|---|
| | YES | NO | SOMETIMES | YES (%) | NO (%) | SOMETIMES (%) | | |
| 1 | 86 | 0 | 0 | 100 | 0 | 0 | 28.6667 | 49.65212 |
| 2 | 77 | 0 | 9 | 89.5 | 0 | 10.5 | 28.6667 | 42.09909 |
| 3 | 57 | 14 | 14 | 66.4 | 16.3 | 16.3 | 28.3333 | 24.82606 |
| 4 | 67 | 14 | 5 | 77.9 | 16.3 | 5.8 | 28.6667 | 33.50124 |
| 5 | 51 | 29 | 6 | 59.3 | 33.7 | 7.0 | 28.6667 | 22.50185 |
| 6 | 47 | 26 | 13 | 54.7 | 30.2 | 15.1 | 28.6667 | 17.15615 |
| 7 | 26 | 45 | 15 | 30.2 | 52.3 | 17.4 | 28.6667 | 15.17674 |
| 8 | 26 | 47 | 13 | 30.2 | 54.7 | 15.1 | 28.6667 | 17.15615 |
| 9 | 52 | 25 | 9 | 60.5 | 29.1 | 10.5 | 28.6667 | 21.73323 |
| 10 | 37 | 37 | 12 | 43.0 | 43 | 14.0 | 28.6667 | 14.43376 |
| 11 | 36 | 33 | 17 | 41.8 | 38.4 | 19.8 | 28.6667 | 10.21437 |
| 12 | 15 | 47 | 24 | 17.4 | 54.7 | 27.9 | 28.6667 | 16.50253 |
| 13 | 45 | 28 | 13 | 52.3 | 32.6 | 15.1 | 28.6667 | 16.01041 |
| 14 | 49 | 28 | 8 | 57.1 | 32.6 | 9.3 | 28.3333 | 20.50203 |
| 15 | 13 | 65 | 6 | 15.4 | 75.6 | 7.0 | 28.0000 | 32.23352 |

The Table 1 also points out that about 66.4% of the respondents considered in this study agrees that the internet is abused while those that are of the opinions that the internet is not abused, or sometimes abused have the same response count of 14 representing 16.3% of the total respondents for each. Over 65% think and believe that the internet is abused and this abuse can be controlled. They tend to be a bit more neutral on whether a specific age range abuses the internet most and whether the freedom of information law and the cybercrime bill in Nigeria have a connection with the internet viewing control. Also over 50% agree that the use of web filters outweigh the risk of using web filters and thereby increasing the chances of controlling the abuse of the internet while indicating strongly that the EFCC have not done enough in combating internet abuse.

## 4.2 System Development

The semantics of the content filter was designed using Unified Modelling Language (UML) specifically the class diagram which is a type of static structure diagram that describes the structure of a system by showing the system's classes, their attributes, operations or methods and the relationships among the classes. The class diagram shown in Fig. 2 illustrates the server acting as the interrelation for the browser and the database check. While Fig. 3 and Fig. 4 shows the program interfaces for testing IP address banned by the Content Filter and for IP address is been removed by the Content Filter from the Database.

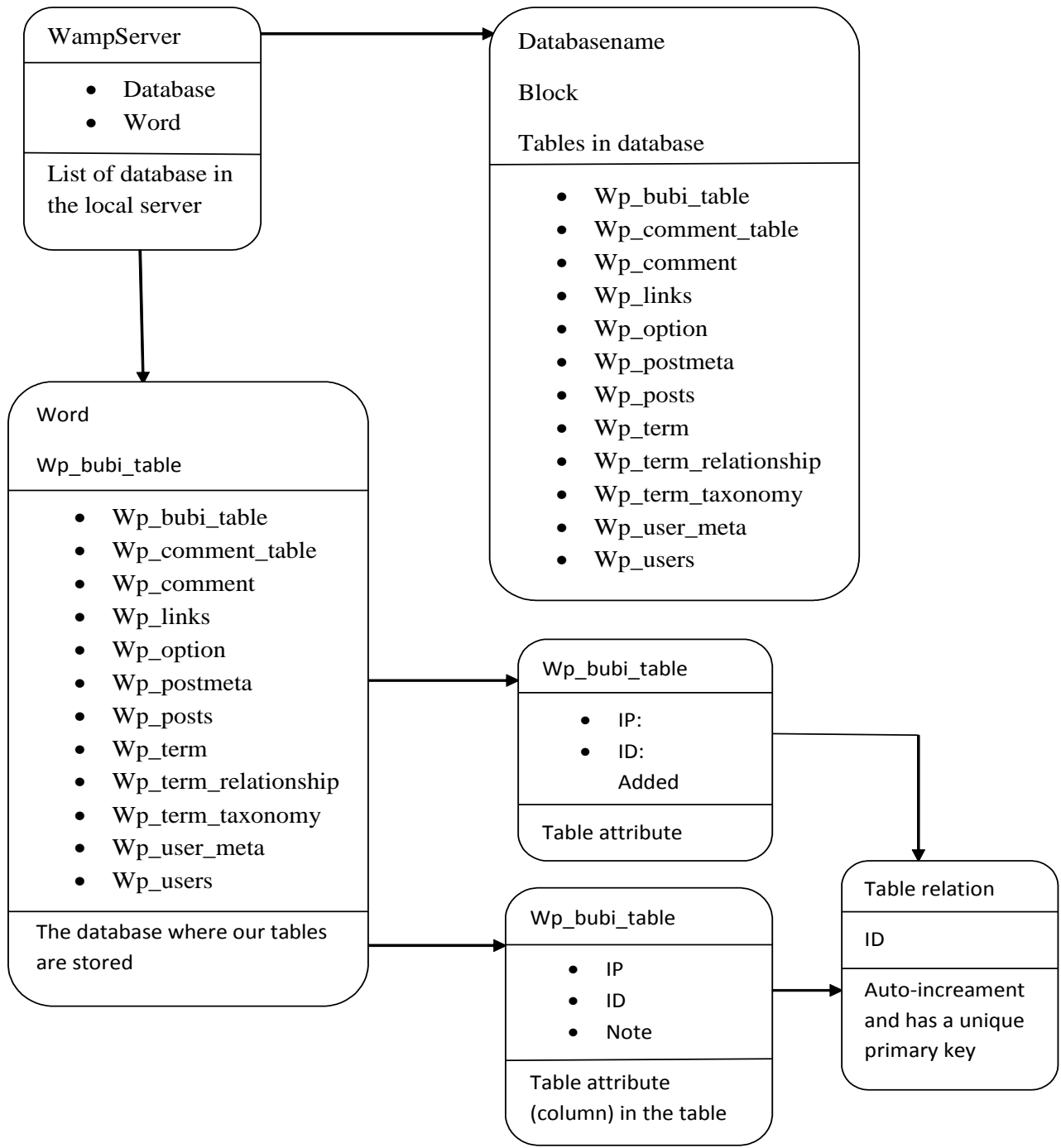

Fig. 2: Proposed System Class Diagram

Fig. 3: Testing IP address banned by the Content Filter

Fig. 4: IP address is been removed by the Content Filter from the Database

Content filters software was developed to ban user from access illegal contents or fraudulent websites. This approach enables the administrator to have access to the client server and any illegal site viewed can be deleted. Since internet viewing by the citizen can now be controlled.

## V. Conclusion and Recommendations

The results of this study reveal that almost all the respondents considered in the survey know about internet and the vast majority of them use it. It was also shown that the internet has been actually abused, and this abuse can be controlled with the government and its agencies such as Economic and Financial Crimes Commission (EFCC) taking the lead. It was suggested that children's access to content on the internet should be restricted while that of the adults should be relaxed. Besides, it was also found out that the freedom of information law and the cybercrime bill in Nigeria have a connection with the internet viewing control. The results also indicated that the EFCC has not done enough in combating internet abuse and more still need to be done.

This study also achieved designing a programme that is capable of filtering illegal websites since each website has a unique IP Address, any website that is illegal can be banned from accessing by typing the IP Address of that particular site. This approach enables the administrator to have an access to the client server and any illegal site viewed can be deleted since internet viewing by the citizen can now be controlled using content filters if not completely eradicated. For instance, if a client enters any website that is fraudulent or illegal in which the IP address is automatically in the database, such website will be banned from being accessed by giving client.

Though the objectives of this project have been achieved, more can be done to improve on the use of content filters to tackle the menace of citizen viewing illegal websites (*internet abuse*). Based on the results of this study, the following recommendations are made:

i.  government should provide more policies that support the use of web filtering on public organizations.

ii. government agencies such as the EFCC should strengthen their efforts in controlling internet abuse

iii. public organizations such as business outfits, academic communities etc should also encourage the use of web content filtering to help fight internet abuse.

iv. the university authorities in various Nigerian institution should affiliate with other online resources centres worldwide so as to enable their student access relevant materials.

**Author's Profiles**

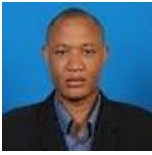

**Shafi'i Muhammad ABDULHAMID** holds M.Sc. degree in Computer Science from Bayero University Kano, Nigeria (2011) and B.Tech. degree in Mathematics/Computer Science from the Federal University of Technology Minna, Nigeria (2004). His current research interests are in Cyber Security, Grid Computing, Cloud Computing, Network Security and Computational Intelligence. He has published many academic papers in reputable journals within Nigeria and Internationally. He is a member of IEEE, International Association of Computer Science and Information Technology (IACSIT), International Association of Engineers (IAENG), The Internet Society (ISOC) and a member of Nigerian Computer Society (NCS). Presently he is a lecturer at the Department of Cyber Security Science, Federal University of Technology Minna and a PhD Scholar at the Faculty of Computing, Universiti Teknologi Malaysia.

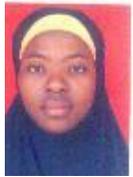

**Fasilat Folagbayo IBRAHIM** obtains a B.Tech. degree in Mathematics/Compter Science from the Federal University of Technology Minna, Nigeria (2012). She is a member of National Association of Computer Science Students (NACOSS) and National Association of Mathematics Students (NAMS). Her current research interests are in Cyber Security, Software Engineering and Computational Intelligence.